\documentclass[twocolumn,showpacs,preprintnumbers,amsmath,amssymb]{revtex4}

\usepackage{graphicx}
\usepackage{dcolumn}
\usepackage{bm}

\DeclareMathOperator{\arcsec}{^{\prime\prime}}
\DeclareMathOperator{\arcmin}{^{\prime}}

\begin{document}

\preprint{arXiv:1008.0426v3 [astro-ph]}

\title{Search for Cosmic Strings in the COSMOS Survey}

\author{J. L. Christiansen}
\email{jlchrist@calpoly.edu}
\affiliation{Department of Physics, California Polytechnic State University, 
San Luis Obispo, California 93407, USA}
\author{J. Goldman}
\affiliation{Department of Physics, 
National University of Singapore, Singapore 117542}
\author{I.P.W. Teng}
\affiliation{Department of Physics, 
National University of Singapore, Singapore 117542}
\author{E. Albin}
\affiliation{Department of Physics, California Polytechnic State University, 
San Luis Obispo, California 93407, USA}
\author{T. Fletcher}
\affiliation{Department of Physics, California Polytechnic State University, 
San Luis Obispo, California 93407, USA}
\author{M. Foley}
\affiliation{Department of Biomedical Engineering, California Polytechnic State University, 
San Luis Obispo, California 93407, USA}
\author{G. F. Smoot}
\affiliation{Lawrence Berkeley National Laboratory, Space Sciences Laboratory 
\\and Department of Physics, University of California, Berkeley, California 94720, USA}



\date{\today}

\begin{abstract}
We search the COSMOS survey 
for pairs of galaxies consistent with the gravitational lensing signature 
of a cosmic string.  The COSMOS survey imaged 1.64 square degrees using 
the Advanced Camera for Surveys (ACS) aboard the Hubble Space Telescope (HST).
Our technique includes estimates of the efficiency for finding the 
lensed galaxy pair. We find no evidence for cosmic strings with a mass 
per unit length of $G\mu/c^2<3.0\times10^{-7}$ out to redshifts greater 
than 0.6 and set 95\% upper limits.  
This corresponds to a global 95\% upper limit of 
$\Omega_{strings}<0.0028$.

\end{abstract}

\pacs{98.80.Cq}
\keywords{Cosmic Strings, gravitational lensing, HST COSMOS survey}

\maketitle

\section{\label{sec:intro}Introduction:}\protect
Cosmic strings are linear topological defects that arise naturally during
symmetry-breaking phase transitions in the early universe 
\cite{kibble-GUT, polchinski-DF}.  
They have also been proposed in string theory models of inflation, occuring 
just after the GUT scale transition \cite{sakellariadou}.
Detailed simulations of the dynamics and interactions of cosmic 
strings predict 
a modern day stochastic network of strings, observable through a variety of 
astrophysical phenomena \cite{hindmarsh, tye}.  
The dimensionless scale of observational interest is  
$G\mu/c^2 \lesssim 10^{-6}$ where $\mu$ is the string energy-density 
\cite{kibble-2}.  

Despite considerable interest within the theory community and multiple proposed
production mechanisms, only a few observations bear on the subject, including 
cosmic microwave background (CMB) \cite{WMAP-3a, WMAP-3, WMAP-5, WMAP-6, jeong}, 
gravitational waves \cite{pulsar, ligo2, ligo-nature}, 
and gravitational lensing \cite{ours, sdss-lensing}.  
The CMB power spectrum shows that cosmic strings are not the dominant 
factor in large-scale structure formation, but that they may contribute up to about
10\% of the observed structure, enough to possibly detect with the Planck data 
\cite{jeong2, PLANCK} 
or other experiments with resolution on small angular scales \cite{CMBpredict}. 

Gravitational lensing of background galaxies by a cosmic string 
is expected to produce a pair of images separated by an angle, 
$\Delta\theta=\delta\sin(\beta)D_{ls}/D_{os}$ where $\delta=8\pi G\mu/c^2$ is
the deficit angle, $D_{ls}$ is the distance between the lensing string ($l$) and
the background source ($s$), $D_{os}$ is the distance between the observer ($o$) 
and the background source, and $\beta$ is the tilt of the string toward the
observer.
  
In our previous search for gravitational lensing by cosmic strings 
in the GOODS survey we concluded with 95\% confidence that
$G\mu/c^2<3.0\times10^{-7}$ out to redshifts greater than 0.5 and 
that $\Omega_{strings}<0.02$ \cite{ours}.
Our aim in this paper is to use the same technique to analyze the 
wider survey carried out by the COSMOS team
with the Hubble Space Telescope (HST) Advanced Camera for Surveys (ACS). 
This survey has a field-of-view that is 24 times larger than the GOODS 
survey and the limits we find are nearly 10 times lower than our previous limits.
Because strings are linear objects and backgrounds scale with the 
field-of-view, we have found it necessary to perform the search in four 
angular bins to improve the signal-to-noise ratio.  

In Section II we describe our data selection and present the correlation 
analysis used to search for cosmic strings in the COSMOS survey.  
We discuss the simulations needed to estimate the signal rates and
detection efficiencies in Section III. These estimates are then used
in Section IV to determine limits on individual
cosmic strings as a function of mass and redshift as well as the global 
limit on the density of cosmic strings.  
Finally, we summarize our results in Sec. V.

\section{\label{sec:data}Data Sample:}\protect
The COSMOS field is 1.64 degrees$^2$ centered 
on RA=10:00:28.6 and DEC=+02:12:21.0. 
Images were taken with the ACS aboard HST between July 2003 and June 2005
\cite{cosmos, cosmos-1}. 
We analyze the publicly available COSMOS Version 1.3 data 
in the F814W (I-band) filter which consists of 81 drizzled tiles with a 
resolution of $0.05\arcsec$/pixel.
We apply a fiducial cut and use the central 1.57 degrees$^2$ of the survey.

\subsection{Source identification}
We use SExtractor version 2.5.0 (Source Extractor) to identify sources in 
the COSMOS survey \cite{SEX-1,SEX} following the {\it Hot} procedure outlined
in \cite{cosmos-cat}.  For shape-sensitive analyses like weak lensing, 
Leauthaud et al. advocate a {\it Hot-Cold} method applied
to un-rotated, undrizzled images.  We 
validate our {\it Hot} catalogs against the public release of the 
well-understood
Leauthaud et al. catalog containing 1.2 million objects which we 
refer to as the {\it LC2} catalog described in \cite{cosmos-cat}.  
We do not perform a catalog-level search using the {\it LC2} catalog 
directly due to the difficulties in estimating the efficiency of 
finding lensed galaxies without access to the original COSMOS fits 
images and processing pipeline.

Several minor modifications need to be applied to Leauthaud's {\it Hot} 
parameters \cite{cosmos-cat} to account for the difference in resolution 
for the publicly available tiles. 
We set the \verb+PIXEL_SCALE+ to 0.5 arcsec, 
use a gaussian filter width of 2.5 pixels,
and set the \verb+DETECT_MINAREA+ to 9 contiguous pixels above threshold.  
We also scale the apertures to 12 pixels, \verb+PHOT_APERTURES+ and \verb+PHOT_AUTOAPERS+.  
Finally, to reduce the deblending slightly, we set the 
\verb+DEBLEND_NTHRESH+ to 32 and the \verb+DEBLEND_MINCONT+ to 0.1.
The resulting catalog contains 812,463 objects with magnitudes brighter than
26.5. 

\subsection{\label{sec:dataGal}Resolved galaxy selection}
\begin{figure}
\includegraphics{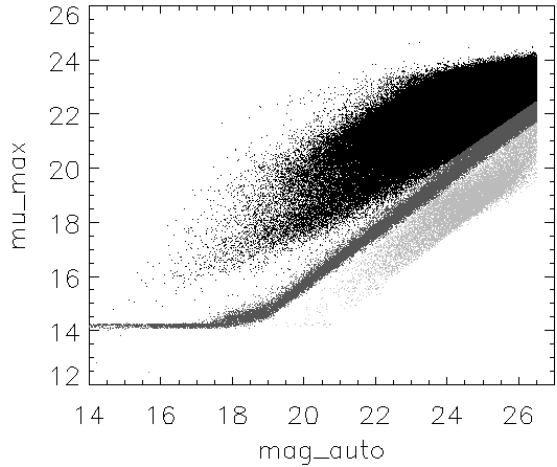}
\caption{\label{fig:stars} 
Resolved galaxies (black points) are a distinct population.  
Also shown are point sources (dark gray) including stars, and spurious objects (light gray)
that are too small to be consistent with the PSF.}
\end{figure}

\protect
We select the resolved galaxies from this sample using the 
correlation of the peak surface brightness,
\verb+MU_MAX+, with the objects magnitude, \verb+MAG_AUTO+.  
Fig.~\ref{fig:stars} shows three regions of interest in the \verb+MU_MAX+ vs 
\verb+MAG_AUTO+ plane: resolved galaxies, point sources
including stars, and spurious detections where the objects are too small to
be consistent with the point spread function.  There are 761,370 resolved 
galaxies, 36,823 point sources, and 14,270 spurious detections.  
The {\it LC2} catalog contains fewer spurious detections due to a prior 
cleaning procedure that included merging of 
small objects with nearby objects from the {\it Cold} catalog 
and removal of objects in regions of elevated noise that occur on the 
borders of the un-rotated tiles as well as near bright stars.  
To reproduce the additional ``by hand'' cleaning, we correlate our resolved 
galaxies with the {\it LC2} catalog.  

We find that 96.4\% of the resolved galaxies with magnitudes brighter than 23$^{rd}$ magnitude 
are also found in the {\it LC2} catalog.  
Our magnitudes generally agree, but have a tail down to smaller values which we 
attribute to over deblending of larger objects.
The 3.6\% of events that don't have a counterpart in the {\it LC2} 
catalog are all found in the diffraction trails of very bright stars or in a few cases 
near the edge of the survey consistent with the more aggressive cleaning.  
The situation is 
understandably slightly worse for dimmer objects; 87.7\% of the galaxies at 
25$^{th}$ magnitude have a counterpart in the {\it LC2} catalog.  
The magnitudes tend to agree well.  The 12.3\% of galaxies that don't have a 
counterpart in the {\it LC2} catalog are again in fiducial regions that were 
removed by hand.  
For the rest of the galaxies, most of them are in regions with 
elevated noise.  
Very few appear to be legitimate detections.

To reproduce the cleaning as much as possible, we remove resolved sources in 
our catalog that are not included in the {\it LC2} catalog.  
This removes about 11.6\% of our resolved galaxies.  
Any inefficiency that comes about from this requirement is included in our 
efficiency estimate.  More importantly, though, this requirement protects us 
from overestimating the cosmic string lensing rate due to spurious detections. 

We post-process our resolved galaxy catalog to
identify the pixels in the image associated with each galaxy.  
First we define a small but encompassing search region about each 
galaxy centroid.  This region is chosen to be three times the size 
of the galaxy 
reported by the catalog coordinates 
(\verb+XMIN_IMAGE+, \verb+YMIN_IMAGE+) and 
(\verb+XMAX_IMAGE+, \verb+YMAX_IMAGE+).
Next we determine the local background characteristics by fitting a 
gaussian to the
small amplitude peak in a histogram of the pixel intensities. 
Finally, we find a bright pixel near the galaxy centroid and iteratively
aggregate neighboring pixels that are $1 \sigma$ above the mean background. 
This process occasionally merges neighboring galaxies.  
In the event that a cluster of pixels reaches the edge of the search region 
or that two galaxies merge, we raise the neighbor threshold to $2 \sigma$ 
above the mean background and repeat the process.  
We continue to raise the threshold until each galaxy is completely contained 
within the search region and does not contain the centroid from any other 
galaxy in the catalog.  
The aim of this procedure is to retain as much unbiased shape 
information as possible.
For 0.15\% of the dimmest sources the threshold is raised so high that 
there are no pixels left in the cluster and we remove these galaxies from 
the sample.  
After selecting resolved galaxies also identified in the {\it LC2} 
catalog and identifying the galaxy pixels, the 
resulting catalogs contain 662,765 resolved galaxies.

\subsection{\label{sec:dataCorr}Matched galaxy pair selection}\protect
The morphological similarity between each pair of galaxies is characterized 
by the correlation and cross-correlation of the two galaxy images \cite{ours}.
We first align the centroids and then calculate the correlation ($CORR$) and 
the cross-correlation ($XCORR$) of the pixel intensities.
\begin{equation}\label{eq:corr}
CORR = \frac{\sum{I_1(x_i,y_i)^2} - \sum{I_2(x_i,y_i)^2}}{\sum{I_1(x_i,y_i)^2} + \sum{I_2(x_i,y_i)^2}}
\end{equation}
\begin{equation}\label{eq:xcorr}
 XCORR = \frac{2\sum{I_1(x_i,y_i)*I_2(x_i,y_i)}}{\sum{I_1(x_i,y_i)^2} + \sum{I_2(x_i,y_i)^2}}  
\end{equation}
where $I(x_i,y_i)$ is the intensity of each pixel in a galaxy and 
the subscript 1 or 2 refers to the galaxies being correlated. 
Galaxies with $CORR$ near a value of zero have very similar magnitudes and 
galaxies with $XCORR$ near a value of one are similar in surface brightness
and shape.
We define matched galaxy pairs as those within the ellipse defined by
\begin{equation}\label{eq:corrcut}
\sqrt{(2*CORR)^2 + (1-XCORR)^2} < 0.29 
\end{equation}
This cut was optimized on simulated lensing events 
and is slightly looser than the one used in our previous GOODS search. 

In this analysis, we consider pairs of galaxies with opening angles, 
$\Delta\theta<15\arcsec$. 
There are 96,413 matched pairs out of 7,081,011 total pairings with 
$\Delta\theta<15\arcsec$ in the COSMOS survey.  
The selected matched pairs are consistent
with the null hypothesis.

\subsection{\label{sec:pairs}Pairs distribution}\protect

\begin{figure}
\includegraphics{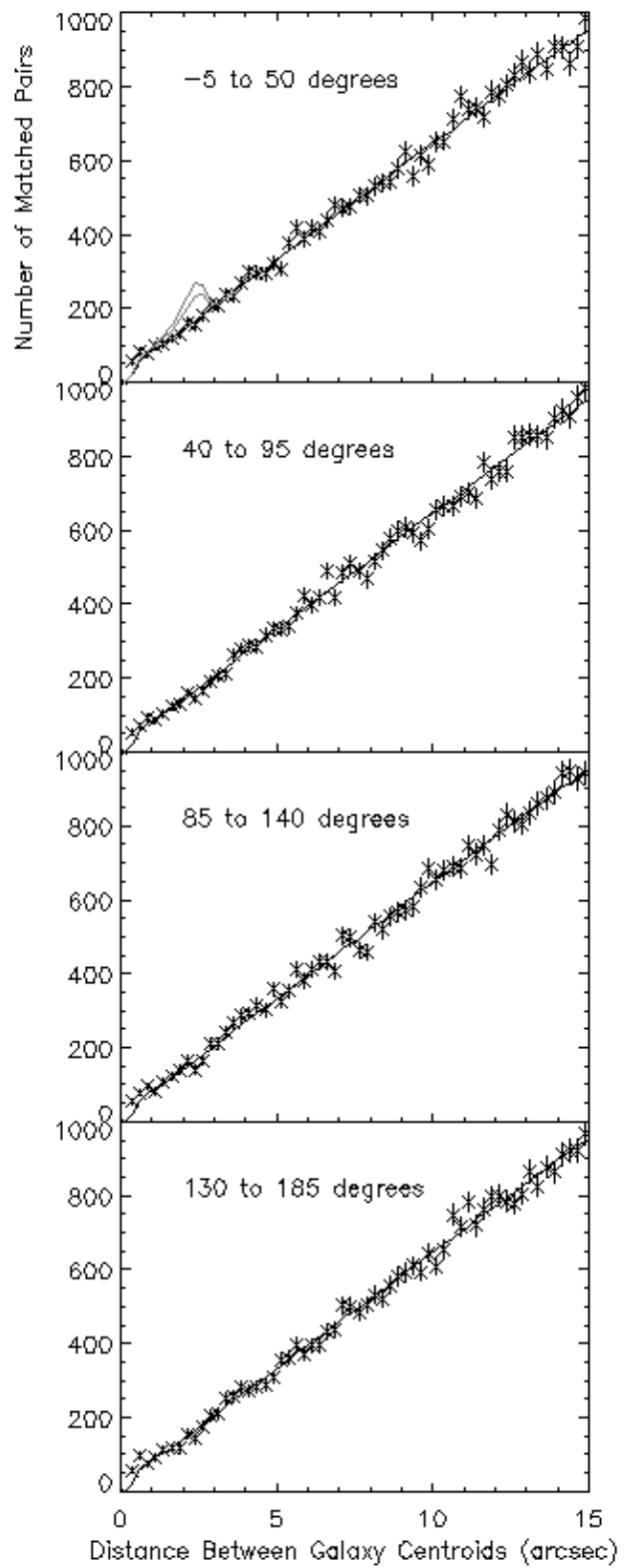}
\caption{\label{fig:pairs}
Pairs of galaxies in four angular bins (points)
are compared to background (solid line).  The top panel shows one example of a simulated 
string (grey).  
The upper simulated string is the total number of pairs expected from the 
simulation with string length of 1.19$^o$, redshift of 0.5, and $\delta\sin\beta$ of $4\arcsec$ .  
The lower simulated string includes measurement inefficiencies.}
\end{figure}


The binned distribution of matched galaxy pairs is shown in 
Fig.~\ref{fig:pairs}.  
The distribution is divided into four overlapping angular 
bins:  -5$^o$:50$^o$, 40$^o$:95$^o$, 85$^o$:140$^o$, and 130$^o$:185$^o$.  
The angular bins correspond to a range of cosmic string angles on the sky.   
The hypothetical string is presumed to pass between the two galaxies in a pair 
at an angle perpendicular to the line connecting the galaxy centroids.  
These bins are needed to reduce the background as the survey gets 
increasingly large.  
Because cosmic strings 
are linear objects, the signal scales with the width of the survey whereas 
the background scales with the area of the survey. 
We do not want to restrict our search to perfectly straight
strings so we keep the number of bins to a minimum.
Overlapping bins are chosen to assure us that we are searching efficiently.   
   
The background shape is characterized by the distribution of all pairs 
of galaxies regardless of size and shape.  
Because strings with masses large enough to create opening angles greater 
than $7\arcsec$ have been ruled out, 
we normalize the background distribution 
to the number of measured matched pairs between  $7\arcsec$ and $15\arcsec$. 
This gives us a reliable estimate of the background at smaller 
opening angles.  
From the background, we observe that SExtractor merges galaxies with opening
angles below $0.4\arcsec$.  

In our signal region, between $0.4\arcsec$ and $7\arcsec$, there are 24 data bins 
and the $\chi^2/24-dof$ of the matched pairs to the background is 
0.99, 1.1, 1.3, and 1.4 for the four angular bins respectively.  The p-values range
from 15\% to 48\%.
Based on the scaled background distribution, we report no 
evidence for an excess of pairs at small opening angles.  
Although there is no statistically significant excess in our signal region, 
we note that there is a small excess of
signal below $1\arcsec$ and that the first data point in the signal region 
is systematically high.
This excess is consistent with the fact that the dimmest galaxies tend to pass 
the correlation cuts more easily than larger galaxies.  
To investigate this region more carefully, we surveyed matched pairs with 
opening angles less than $1\arcsec$ by hand.  We find that 
matched pairs tend to be discovered in regions of high galaxy density and 
that these regions tend to have other lensing candidates at the fairly high 
rate of $0.5/1\arcmin$.  A full statistical 
analysis of these pairs shows no evidence of lensing and gives us a 
calibration curve for the background in high density regions.

\section{\label{sec:sim}Cosmic String Simulations:}
\begin{figure}
\includegraphics{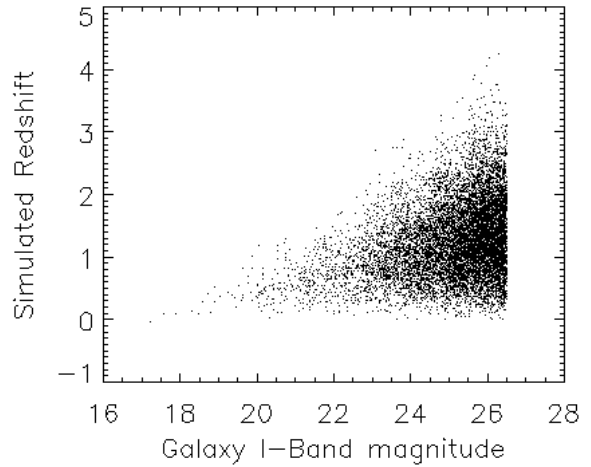}
\caption{\label{fig:redshift}
The simulated redshift distribution for galaxies in a typical COSMOS tile. }
\end{figure}

\protect
Simulations are used to estimate both the number of lensed galaxy pairs 
based on the local density of galaxies in the COSMOS survey as well as the 
efficiency of finding those pairs.
The simulation of lensed galaxy pairs is the same as used previously in 
our GOODS search \cite{ours}.  The idea is to use the density of galaxies
in our catalog of sources to monte carlo the number of pairs that
would exist from any theoretical string crossing our survey.  
An important aspect of this calculation is the simulation of the galaxy 
redshifts.  
We use the same parameterization of the redshift distribution as \cite{massey}.
For these studies, COSMOS tile 55 is used as a typical example.  
The simulated redshift distribution of the galaxies in this tile is shown in
Fig. ~\ref{fig:redshift}.

The advantage
of a catalog-level signal simulation is that we can quickly embed as many 
strings
as needed to get an accurate estimate of the average number of lensed galaxies
observed from a particular set of string parameters, including redshift and 
$\delta\sin\beta$.  
An example of the result from this simulation is shown
in the top panel of Fig.~\ref{fig:pairs}.  

\begin{figure}[b]
\includegraphics{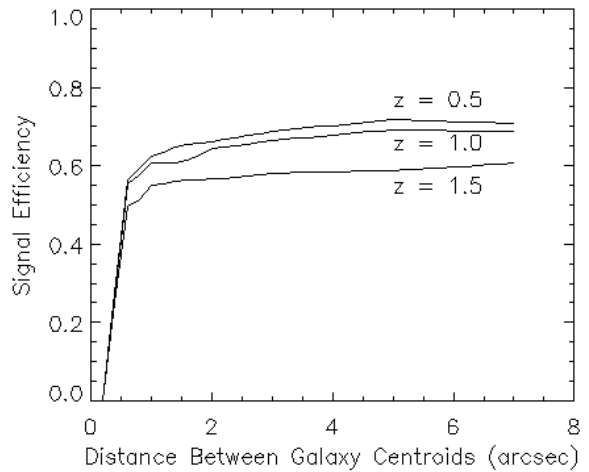}
\caption{\label{fig:eff}
Efficiency of detecting pairs of galaxies lensed by a cosmic string as a 
function of pair opening angle and redshift.}
\end{figure}

Fig.~\ref{fig:eff} summarizes the efficiencies as a function of the
opening angle between the galaxies and the string redshift.  These efficiencies
were estimated by embedding galaxies as realistically as possible into COSMOS 
tiles and then processing the modified tiles back through our analysis chain 
to see 
how many embedded sources are found.  The curves include both the efficiency 
of identifying the embedded galaxies with SExtractor and the correlation and 
cross 
correlation selection cuts. 
Below $0.4\arcsec$ galaxies are merged by SExtractor and the pair is lost. 
For dim galaxies, which tend to have higher redshifts, 
noise in the galaxy detection becomes an increasingly important effect.

\section{\label{sec:results}Results:}\protect
The distribution of matched galaxy pairs shown in Fig.~\ref{fig:pairs} 
rises nearly linearly.  
This is expected for the background pairs and any excess above the background
could be evidence of a cosmic string.
For comparison, simulated pairs from a cosmic string at a redshift of 0.5 and 
$\delta\sin\beta$ of $4\arcsec$ are included in the top panel.  
They are normalized to the mean length of a string crossing the 
survey, $1.19^o$.
The upper curve is the total simulated signal.  The lower curve includes 
the detection inefficiencies from Fig.~\ref{fig:eff}.  

\begin{figure}[t]
\includegraphics{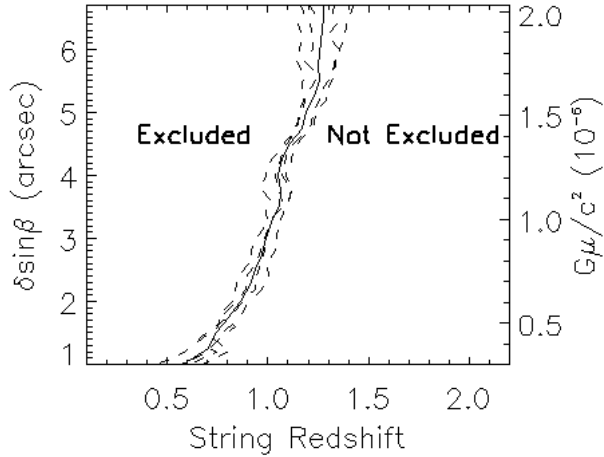}
\caption{\label{fig:limits} 
95\% upper limits for lensed galaxies produced by a cosmic string as
a function of the string mass and redshift.  Dashed lines are the individual limits
from each angular bin shown above.  The solid line is the average limit for all directions.}
\end{figure}

We compare a wide variety of predicted cosmic string signals to the data to
determine limits. Signal pairs, $n_s$, are summed from $0.4\arcsec$ to 
a maximum opening angle beyond which there is no more signal.  
The observed matched pairs, $n_{obs}$, and the background, $n_{bkg}$, 
from Fig.~\ref{fig:pairs} are summed over the same range of opening angles.
We then compute the single-sided Neyman 95\% confidence limit, $n_{lim}$, which
is the minimum signal that is consistent with background fluctuations.  
Signals with $n_s>n_{lim}$ are excluded by the data.
The resulting 95\% upper limits are shown in Fig.~\ref{fig:limits}.  
All four angular bins yield similar limits that 
extend from $1\arcsec<\delta\sin\beta<7\arcsec$.  
We average the four angular bins for the global limit.
Taking the mean tilt of a string with respect to the observer to be 
$<\sin\beta> = 2/\pi$ we relate the opening angle to the 
mass scale via the factor
$8\pi\frac{G\mu}{c^2} = \delta<\sin\beta>$ shown on the right-hand axis.
We see no evidence for cosmic strings out to a redshift greater than 
0.5 and place a 95\% upper limit of $G\mu/c^2<3.0\times10^{-7}$.

\begin{figure}[t]
\includegraphics{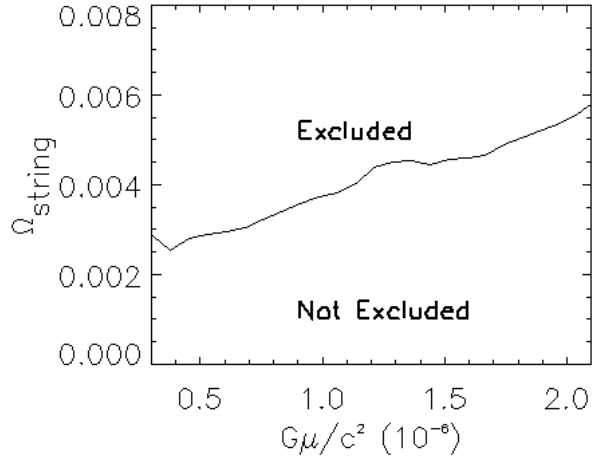}
\caption{\label{fig:omega}
95\% upper limits on the global parameter, $\Omega_{strings}$, as a 
function of the string mass.}
\end{figure}

If strings are rare occurrences, it is likely that none would appear in the 
COSMOS field (fov = 1.57 degrees$^2$) and that other surveys may yield 
different results.  
Because strings are line-like objects, we expect that the COSMOS field of 
view would intersect a single string about 0.5\% of the time.
Using a simple geometric monte-carlo we determine that approximately 
820 straight strings randomly placed in a volume of comoving radius, $\eta$,
are needed for a 95\% detection rate in a COSMOS fov.  
In this case, the average COSMOS fov contains 3 detectable 
string crossings as expected from Poisson statistics.
This string density corresponds to a total length of string, 
$L_{tot} = 1200\eta$, in the volume and we use the invariance 
of $L_{tot}/\eta$ to set limits on $\Omega_{strings}$.
\begin{equation}\label{eq:omega}
\Omega_{strings} = \frac{\mu(L_{tot}/\eta)\eta}{(4/3)\pi\eta^3}
                   \times\frac{8\pi G}{3H_0^2}
\end{equation}
where $\eta$ is the comoving distance 
computed with $h=0.7$, $\Omega_M=0.27$, and $\Omega_{\Lambda}=0.73$. 
Fig.~\ref{fig:omega} shows the string densities excluded by this method.
The limit excludes a string density that is 0.28\% of the critical density 
for the smallest mass strings and rises to 0.4\% for more massive strings.  
Each limit assumes that all cosmic strings have the same $G\mu/c^2$ as 
predicted by some models.  Because non-uniform string networks are 
predicted to have a very steep spectral index, (-10$<$index$<$-6) \cite{tye}, 
strings at the measurement threshold will dominate the sensitivity.  
  
A wide variety of theoretical models predict 
$\Gamma = \Omega_{strings}/(8\pi G\mu)$ of order 10 \cite{tye}.
Our measurement is consistent with $\Gamma=375$ for low-mass strings 
and $\Gamma=110$ for our largest-mass strings and thus indicates that
we are not yet sensitive to masses of current theoretical interest.

\section{\label{sec:conclusion}Conclusion:}\protect
We have used the COSMOS survey field to search for cosmic strings.  
We find no evidence for the gravitational lensing signature.  
We have included the observational efficiencies in our analysis using 
the same technique we used previously on the GOODS survey. 
Figs.~\ref{fig:limits} and \ref{fig:omega} summarize our results.
We set 95\% upper limits of $G\mu/c^2<3.0\times10^{-7}$ 
out to redshifts greater than 0.6 which leads in turn to a global limit of
$\Omega_{strings}<0.0028$.
We note that, while these results have as their foundation the hypothesis
of long straight strings, they also exclude strings with moderate curvature.

The global limit on $\Omega_{strings}$ is nearly 10 times stronger 
than our previously published limit \cite{ours}.  
We want to emphasize that our technique is complementary to other methods.
We excluded masses that are smaller than those excluded by other direct
CMB searches \cite{jeong}.  However, our masses are larger than those 
reported by parameter fits to the CMB \cite{WMAP-3a, WMAP-3, WMAP-5, WMAP-6} 
as well as gravitational wave searches \cite{pulsar, ligo2}.  
The recent microlensing search carried out on SDSS quasars 
\cite{sdss-lensing} is sensitive to very low-mass strings, but
reports a similar sensitivity for $\Omega_{strings}$.
Our direct searches may be sensitive to a particular class of 
strings that could be missed by other searches.

\section{\label{sec:ack}Acknowledgements:}\protect
We would like to thank Kevin James for his early
participation in this analysis project.  We also thank 
Alexie Leauthaud for useful discussions of the COSMOS dataset 
and her catalog.
This research used resources of the National Energy Research Scientific 
Computing Center, which is supported by the Office of Science of the 
U.S. Department of Energy under Contract No. DE-AC02-05CH11231. 


\bibliography{../cosmos/cosmos_string}

\end{document}